\documentclass[fleqn,twoside]{article}
\usepackage{espcrc2,colordvi,psfig}
\newcommand{\fillrhd}{{\rhd\kern-.81em\bullet}\kern-.47em\triangleright}

\title{Higher Loop Results for the Plaquette, Using the Clover and
Overlap Actions\thanks{Presented by A. Athenodorou}}

\author{A. Athenodorou\address{Department of Physics, University of
        Cyprus, Nicosia CY-1678, Cyprus}, H. Panagopoulos$^{\rm a}$,
        A. Tsapalis$^{\rm a}$} 

\begin{document}

\begin{abstract}
We calculate the perturbative value of the free energy in QCD on the
lattice. This quantity is directly related to the average plaquette. 

Our calculation is done to 3 loops using the clover action for
fermions; the results are presented for arbitrary values of the clover
coefficient, and for a wide range of fermion masses.

In addition, we calculate the 2 loop result for the same quantity,
using the overlap action.

\end{abstract}

\maketitle
\section{INTRODUCTION}

In this work, we compute the perturbative expansion of the
average plaquette, in \Blue{$SU(N)$} gauge theory with \Blue{$N_f$} fermion
flavours.
We present separate calculations using the
\Blue{clover} action (\Blue{3-loops}), and the \Blue{overlap} action
(\Blue{2-loops}) \cite{APT}. The simpler case of Wilson fermions was performed in \cite{AFP}.

The average plaquette can be related to the
perturbative free energy of lattice QCD, as well as to the expectation
value of the action.

The results can be used:
a) In improved scaling schemes, using an appropriately defined
effective coupling.
b) In long standing efforts, starting with \cite{DR}, to determine
the value of the gluon condensate.
c) In studies of the interquark potential \cite{BB}.
d) As a test of perturbation theory, at its limits of applicability.

In standard notation, the action consists of a gluonic and a fermionic
part::
\begin{eqnarray}
  S &=& S_G + S_F , \nonumber \\
 S_G &=& \beta \sum_P E_G(P)
\end{eqnarray}
with
\begin{equation}
 E_G(P) =  1 - \hbox{Re Tr} (P) / N
\end{equation}
The sum runs over all $1{\times}1$ plaquettes $P$. The fermionic
action $S_F$, in the clover case, contains the Wilson term with bare
fermionic mass \Blue{$m$}, and the standard clover term multiplied by
the coefficient \Blue{$c_{\rm SW}$}, which is a \Blue{free} parameter in the present
work; it is normally 
tuned in a way as to minimize ${\cal O}(a)$ effects.

The average value of the action density, $S/V$, is directly related to
the average plaquette; in particular, for the gluonic part we have: 
\begin{equation}
\langle S_G/V \rangle = 6 \,\beta\,\langle E_G(P)\rangle.
\end{equation} 
As for $\langle S_F/V\rangle$, it is trivial in any action which is
bilinear in the fermion fields~\cite{AFP}, and leads to:
\begin{equation}
\langle S_F/V\rangle = - 4 N N_f
\label{ef}
\end{equation}

We will calculate $\langle E_G\rangle$ in perturbation theory:
\begin{equation}
 \Blue{\langle E_G \rangle} = \Blue{c_1} \; g^2 + \Blue{c_2} \; g^4 + \Blue{c_3} \; g^6 + \cdots
\label{expansion}
\end{equation}
The $n$-loop coefficient can be written as $c_n = c^G_n + c^F_n$ where
$c^G_n$ is the contribution of diagrams without fermion loops and
$c_n^F$ comes from diagrams containing fermions. The coefficients
$c^G_n$ have been known for some time up to 3 
loops~\cite{ACFP,AFP}. The coefficients $c_n^F$ are also known to 3
loops for Wilson fermions~\cite{AFP}; in the present work we extend
this computation to clover fermions. 

The calculation of $c_n$ proceeds most conveniently by computing first the free energy $-(\ln Z)/V$, where $Z$ is the full partition function
\begin{equation}
Z \equiv \int [{\cal D}U {\cal D}\bar\psi_i {\cal D}\psi_i] \exp(-S) .
\label{Z}
\end{equation}
The average of $E_G$ is then extracted as follows
\begin{equation}
\langle E_G \rangle = - {1 \over 6}\, {\partial \over {\partial \beta}}\, \left( {\ln Z \over V} \right) .
\label{e}
\end{equation}
In particular, the perturbative expansion of $(\ln Z)/V$ :
\begin{equation}
(\ln Z)/V = d_0 -{3 (N^2{-}1)\over 2}\,\ln\beta + {d_1\over\beta} + {d_2\over\beta^2} + \cdots
\end{equation}
leads immediately to the relations: \\ $c_2= d_1/(24N^2)$, $c_3= d_2/(24N^3)$.

\section{CLOVER FERMIONS}

Up to 3 loops, a total of 62 diagrams contribute, of which 26 involve fermionic
loops. The involved algebra of the lattice 
perturbation theory was carried out using our computer package in Mathematica.
The value for each diagram is computed numerically for a
sequence of finite lattice sizes.
Diagrams must be grouped in several infrared-finite
sets, before extrapolating their values to infinite lattice size;
extrapolation leads to a (small) systematic error, which 
is estimated quite accurately.

The pure gluonic contributions are already known:
\begin{eqnarray}
 c_1^G &=& {{N^2 {-} 1} \over 8 \;N} , \\
 c_2^G &=& \left( N^2 {-} 1 \right) (0.0051069297 -
           {1 \over {128 \; N^2}}) , \nonumber \\
 c_3^G &=& \left( N^2 {-} 1 \right) ( {0.0023152583(50) \over N^3}
 \nonumber \\
& & - {0.002265487(17) \over N} + 0.000794223(19) \; N ) . \nonumber 
\end{eqnarray}

Fermionic contributions take the form:
\begin{eqnarray}
 c_1^F &=& 0  \; ,\nonumber \\
 c_2^F &=&(N^2 {-} 1) \Blue{h_2} \; {{ N_f} \over N} , \\
 c_3^F &=& \left(N^2 {-} 1\right)
          ( \Blue{h_{30}} \; N_f + \Blue{h_{31}} \; {N_f \over N^2} + 
           \Blue{h_{32}} \; {N_f^2 \over N}) . \nonumber
\end{eqnarray}
The coefficients \Blue{$h_2, h_{30}, h_{31}, h_{32}$} depend polynomially on
the clover parameter $c_{\rm SW}$:
\begin{equation}
 \Blue{h_2} {=} h_2^{(0)} {+} h_2^{(1)} \, c_{\rm SW} {+} h_2^{(2)} \, c_{\rm SW}^2 
\end{equation}
$$ \Blue{h_{3i}} {=} h_{3i}^{(0)} {+} h_{3i}^{(1)} \, c_{\rm SW} {+}
 h_{3i}^{(2)} 
 \, c_{\rm SW}^2 {+} h_{3i}^{(3)} \, c_{\rm SW}^3 {+} h_{3i}^{(4)} \,
 c_{\rm SW}^4 $$
We have calculated the values of \Blue{$h_2^{(j)}$}, \Blue{$h_{3i}^{(j)}$} 
for typical values of the bare mass \Blue{$m$} (see Ref. \cite{APT}
for detailed numerical values).

We list below some typical examples of values for $\langle
E_G\rangle$, setting $N=3$. For $N_f=0$ we have:
\begin{equation}
(1/3)\, g^2 + 0.0339109931(3) \, g^4
+ 0.0137063(2) \, g^6
\end{equation}
For $N_f=2$ and
$m=-0.518106$ (corresponding to $\kappa = (8+2m)^{-1} = 0.1436$):
\begin{equation}
\begin{array}{lll}
c_{\rm SW} = 0.0 :\ &(1/3)\, g^2 &+ 0.026185200(3) g^4 \\
&&+ 0.0119649(3) g^6, \\[0.5ex]
c_{\rm SW} = 2.0 :\ &(1/3)\, g^2 &+ 0.013663456(3) g^4 \\
&&+ 0.0110200(13) g^6.
\end{array}
\end{equation}
For $N_f=2$ and $m=0.038$:
\begin{equation}
\begin{array}{lll}
c_{\rm SW} = 0.0 :\  &(1/3)\, g^2 &+ 0.030438866(3) \ g^4 \\
&&+ 0.0138181(2) \ g^6, \\[0.5ex]
c_{\rm SW} = 1.3 :\  &(1/3)\, g^2 &+ 0.025219798(9) \ g^4 \\
&&+ 0.0129659(5) \ g^6, \\[0.5ex]
c_{\rm SW} = 2.0 :\  &(1/3)\, g^2 &+ 0.01800170(1)  \ g^4 \\
&&+ 0.012948(1) \ g^6, \\[0.5ex]
\end{array}
\end{equation}
For $N_f=3$ and $m=0.038$:
\begin{equation}
\begin{array}{llll}
c_{\rm SW} = 0.0 :\  &(1/3)\, g^2 &+ 0.028702803(5) \ g^4 \\
&&+ 0.0139032(2) \ g^6, \\[0.5ex]
c_{\rm SW} = 1.3 :\  &(1/3)\, g^2 &+ 0.02087420(1) \ g^4 \\
&&+ 0.0128495(8) \ g^6, \\[0.5ex] 
c_{\rm SW} = 2.0 :\  &(1/3)\, g^2 &+ 0.01004706(2) \ g^4 \\
&&+ 0.013547(1) \ g^6. \\[0.5ex]
\end{array}
\end{equation}

\section{OVERLAP FERMIONS}

The fermionic action now reads \cite{Neuberger}:
\begin{equation}
S_f =  \sum_f \sum_{x,y} \bar{\psi}^f_xD_{\rm N}(x,y)\psi^f_y.
\label{latact}
\end{equation}
with: $D_{\rm N} = M_0 \left[1 + X (X^\dagger X)^{-1/2} \right]$,
and: $X = D_{\rm W} - M_0$; the sum on $f$ runs over all
flavors. Here, $D_{\rm W}$ is the massless 
Wilson-Dirac operator with $r=1$, and $M_0$ is a free parameter whose
value must be in the range $0 < M_0 < 2$, in order to guarantee the
correct pole structure of $D_{\rm N}$.

Fermionic vertices are obtained by separating the Fourier
transform of $D_N$ into a free part (inverse propagator $D_0$) and an
interaction part $\Sigma$ \cite{KY,AFPV}:
\begin{equation}
{1\over M_0}\, D_{\rm N}(q,p) {=} D_0(p) (2\pi)^4\delta^4(q-p) + \Sigma(q,p)
\end{equation}
\begin{equation}
\Sigma(q,p) {=}
{1\over \omega(p) + \omega(q)}
[X_1(q,p) \end{equation}
\[\qquad - {1\over \omega(p)\omega(q)} X_0(q) X^\dagger_1(q,p) X_0(p)]\]
\[\qquad +{1\over \omega(p) + \omega(q)}
[X_2(q,p) \]
\[\qquad - {1\over \omega(p)\omega(q)} X_0(q) X^\dagger_2(q,p)
X_0(p)] \]
\[+\int {d^4 k\over (2\pi)^4}
{1\over \omega(p) + \omega(q)}
{1\over \omega(p) + \omega(k)}
{1\over \omega(q) + \omega(k)}\times\]
\[\qquad\Bigl[
-X_0(q)X_1^\dagger(q,k)X_1(k,p) \]
\[\qquad\ -X_1(q,k)X_0^\dagger(k)X_1(k,p)\]
\[\qquad\ -X_1(q,k)X_1^\dagger(k,p)X_0(p) \]
\[\qquad\ +{\omega(p)+\omega(q)+\omega(k)\over \omega(p)\omega(q)\omega(k)}\times\]
\[\qquad\quad X_0(q)X_1^\dagger(q,k)X_0(k)X_1^\dagger(k,p)X_0(p)\Bigr] + {\cal{O}}(g^3) \]

\noindent
where $X_0, X_1, X_2$ denote the parts of the Dirac-Wilson operator
with 0, 1, 2 gluons (of order ${\cal{O}}(g^0)$, ${\cal{O}}(g^1)$,
${\cal{O}}(g^2)$, respectively), and:
\begin{equation}
\omega(p) {=} ( \sum_\mu \sin^2 p_\mu + [ 
\sum_\mu (1-\cos p_\mu ) - M_0]^2)^{1/2}
\end{equation}

From the form of $\Sigma(q,p)$, we see that the
2-gluon vertex splits into two different types of contributions:
\centerline{\psfig{width=3truecm,file=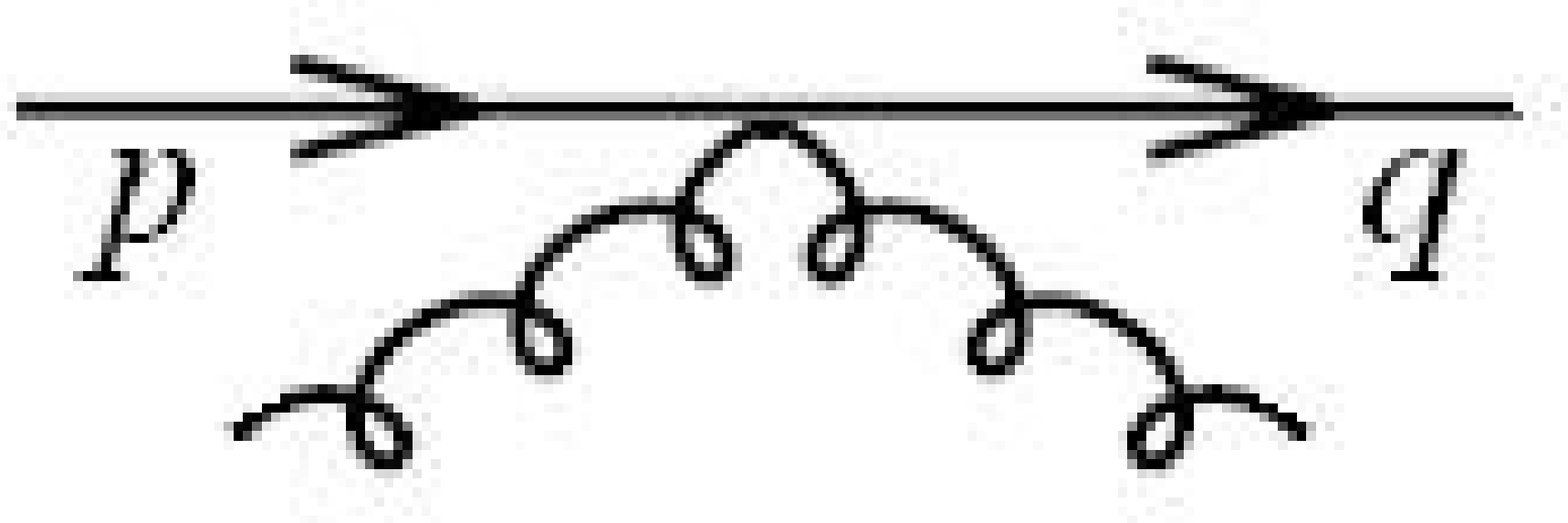}\quad
 \qquad\psfig{width=3truecm,file=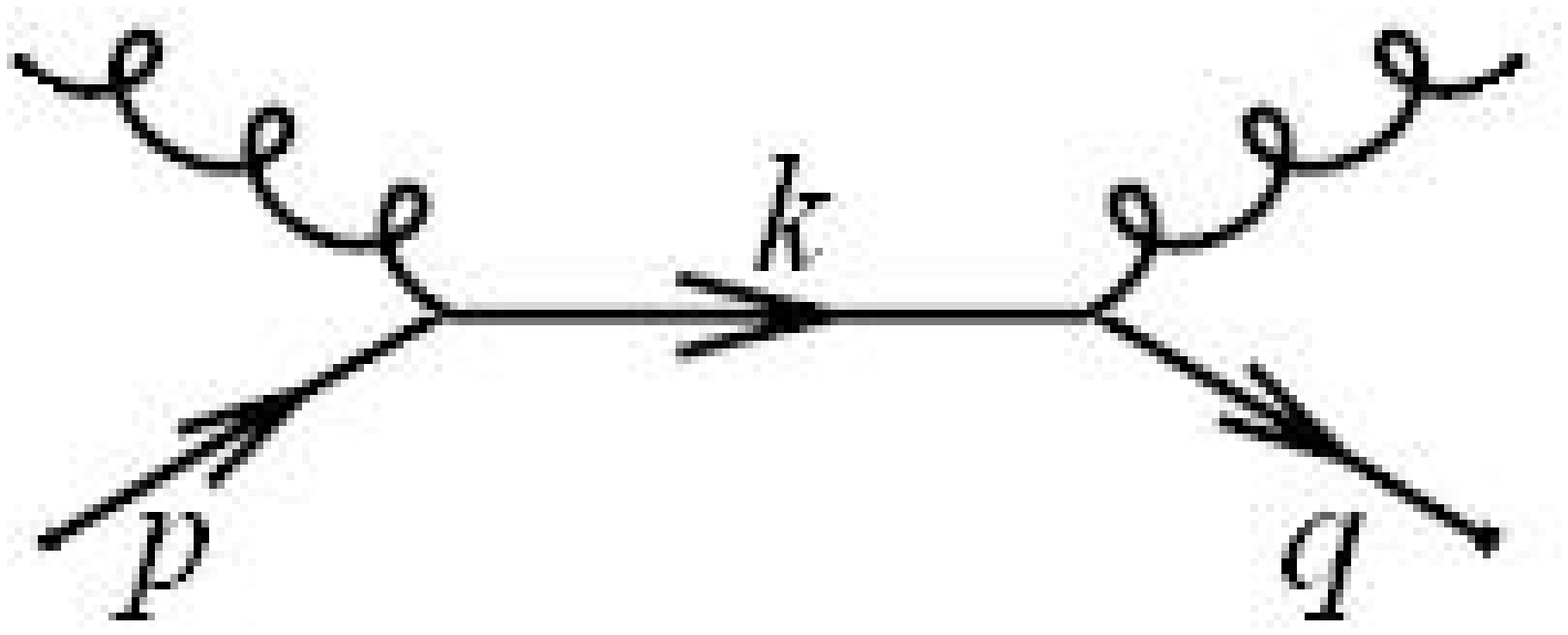}}

Only two fermionic diagrams contribute to 2-loops, leading to: 
\begin{equation}
c_1^F = 0, \qquad c_2^F =\left(N^2 - 1\right) \Blue{h_2} \; {{ N_f}
\over 12\,N} 
\end{equation}
The quantity $\Blue{h_2}$ now depends only on the parameter $M_0$ ; our
result is shown in the graph below.
\centerline{\psfig{width=7truecm,file=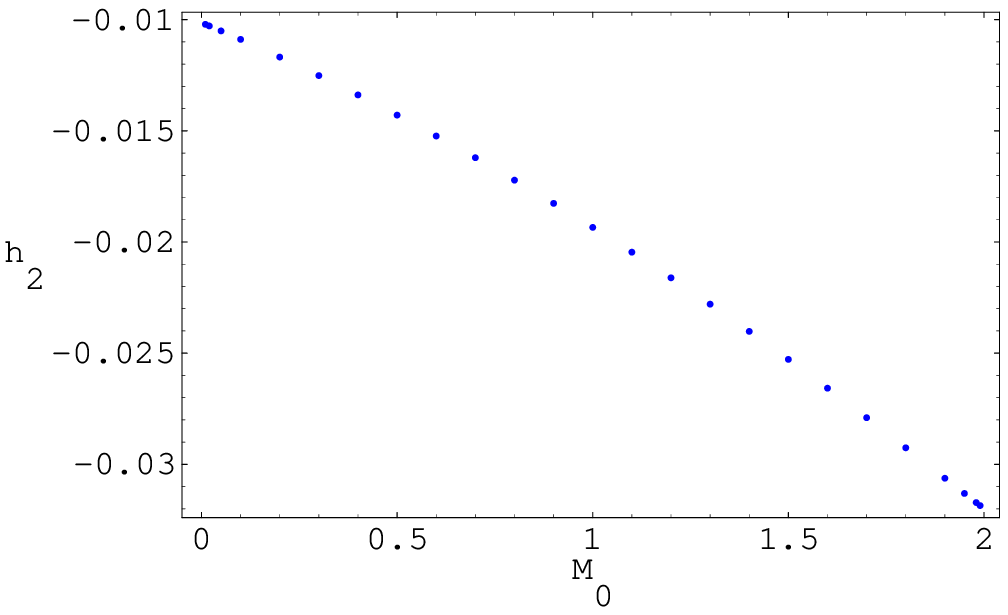}}
It is worth noting that the dependence on $M_0$ is
smooth all the way to the endpoint values $M_0 = 0, 2$, despite the
change in propagator poles at these values.

Typical values of $\langle E_W\rangle$\,: Setting $N=3$,
$N_f=2$, we obtain:

\begin{equation}
\begin{array}{lll}
M_0 = 0.01 : &(1/3)\, g^2 &+ 0.029372693(2)\, g^4, \\[0.5ex]
M_0 = 1.99 : &(1/3)\, g^2 &+ 0.01975396(10)\, g^4, \\[0.5ex]
N_f = 0 :        &(1/3)\, g^2 &+  0.03391099316\, g^4 .
\end{array}
\end{equation}

\end{document}